\documentclass[aps,pra,superscriptaddress,twocolumn]{revtex4}
\usepackage{graphicx}
\usepackage{epstopdf}
\usepackage{tabularx}
\usepackage{amssymb}
\usepackage{amsmath}
\usepackage{bm}
\usepackage{graphpap}
\usepackage{multirow}
\usepackage{notes2bib}
\usepackage{color}
\usepackage[version=3]{mhchem} 
\begin{document}
\title{First-principles many-body study of the electronic and optical properties of \ce{CsK2Sb}, a semiconducting material for ultra-bright electron sources}
\author{Caterina \surname{Cocchi}}
\email{caterina.cocchi@physik.hu-berlin.de}
\affiliation{Physics Department, Humboldt-Universit\"at zu Berlin, Berlin, Germany}
\affiliation{IRIS Adlershof, Humboldt-Universit\"at zu Berlin, Berlin, Germany}
\author{Sonal \surname{Mistry}}
\affiliation{Helmholtz-Zentrum Berlin, Germany}
\author{Martin \surname{Schmei{\ss}er}}
\affiliation{Helmholtz-Zentrum Berlin, Germany}
\author{Julius \surname{K\"uhn}}
\affiliation{Helmholtz-Zentrum Berlin, Germany}
\author{Thorsten \surname{Kamps}}
\affiliation{Helmholtz-Zentrum Berlin, Germany}
\affiliation{Physics Department, Humboldt-Universit\"at zu Berlin, Berlin, Germany}
%
%

\begin{abstract}
We present a comprehensive first-principles investigation of the electronic and optical properties of \ce{CsK2Sb}, a semiconducting material for ultra-bright electron sources for particle accelerators. 
Our study, based on density-fuctional theory and many-body perturbation theory, provides all the ingredients to model the emission of this material as a photocathode, including band gap, band dispersion, and optical absorption.
An accurate description of these properties beyond the mean-field picture is relevant to take into account many-body effects.
We discuss our results in the context of state-of-the-art electron sources for particle accelerators to set the stage towards improved modeling of quantum efficiency, intrinsic emittance, and other relevant quantities determining the macroscopic characteristics of photocathodes for ultra-bright beams.
\end{abstract}

\maketitle
\section{Introduction}

First-principles methodologies for electronic structure calculations have experienced impressive advances in the last few decades.
While density-functional theory (DFT)~\cite{hohe-kohn64pr,kohn-sham65pr} remains the flagship method for \textit{ab initio} modeling of ground-state properties, the state-of-the-art formalism for describing electronic and optical excitations is currently many-body perturbation theory (MBPT) based on Green's function approaches~\cite{onid+02rmp}.
Band structures including the quasi-particle (QP) correction are obtained from the $GW$ scheme~\cite{hedi65pr,hybe-loui85prl,hybe-loui86prb} while absorption spectra, including excitonic effects, are computed from the solution of the Bethe-Salpeter equation (BSE)~\cite{stri88rnc,hank-sham80prb,rohl-loui00prb}.
Thanks to the quick progress and the increasing power of the available computational resources and infrastructures, MBPT, which is significantly more expensive than DFT~\cite{onid+02rmp}, can now be routinely applied to realistic systems. 
This opportunity has produced a strong impact on many scientific and technological areas, including, as prominent examples, the development of solar cells~\cite{dean14acr,bren+16natrm,oba-kuma18apex} and materials for energy storage~\cite{jain+16natrm,yoon+17afm}.

In light of these successful applications, it is natural to expect potential benefits from such advanced first-principles studies also for other tasks.
For example, in the field of particle accelerator physics several groups around the globe are striving for photocathode growth optimization in view of enhancing the performance of semiconducting materials as electron sources for ultra-bright beams~\cite{dowe+10nimpra,musu+18nimpra}.
Among them, it is worth mentioning the high brightness electron beam group at the Helmoltz-Zentrum Berlin working in the framework of the bERLinPro project ~\cite{abo+17proceeding,neum+18ipac}.
In this context, a hot topic concerns the optimal growth procedures of bi-alkali photocathodes, which have been reaching their maturity both as detectors and electron sources~\cite{dowe+93apl}.
The main challenge lies in the reproducible growth of photocathodes with high quantum efficiency (QE), low transverse emittance, and long operational lifetime. 
Along with this issue, another open question to address urgently is the description of photoemission, which is the dominant emission mechanism considered for bright electron beam generation. 
In a typical photoelectron source, electrons are emitted by the photocathode, which is located inside an accelerating gap and illuminated by short laser pulses, so that the transverse and longitudinal shape and size can be controlled on a picosecond to femtosecond time scale via the drive laser pulse. 
The emitted electrons are rapidly accelerated to relativistic energies, thus partially mitigating the brightness deteriorating effects from space charge forces.
The need to overcome the current technology largely based on metallic samples is dictated by the need to improve the yield of emittance and the QE of the photocathode in order to achieve the high repetition rates that are required by free electron lasers, ultra-fast diffraction and microscopy, inverse Compton scattering, and accelerator driven THz radiation sources~\cite{bouc+09nimpra,musu+10rsi,musu+18nimpra}.
To this end, insight from first-principles calculations is considered essential for the microscopic description of the electronic structure of these materials, which is the principal ingredient to model photoemission also on the macroscopic scale.
However, it is reasonable to expect that the mean-field framework of DFT is not suitable for a quantitative representation of the complexity of the photoemission process in semiconductors.
Above all, the determination of the band gap in these systems requires the inclusion of electron-electron correlation as provided by the $GW$ approach of MBPT. 
Likewise, electron-hole interactions may play a crucial role when the material absorbs a photon, due to the relative low screening compared to metals. 
Also electron-phonon scattering mechanisms, representing the dominant processes in the diffusion of photo-excited electrons towards the surface, should be additionally taken into account.

In this paper, we present a comprehensive study of the electronic and optical properties of \ce{CsK2Sb}, an exemplary crystalline material belonging to the family of multi-alkali antimonides, which are largely adopted as photocathodes~\cite{smed+09aipcp,schu+13aplm,musu+18nimpra}.
Along with the detailed characterization of this system, the scope of this work is to demonstrate the capability of first-principles many-body methods to give insight into the electronic structure and excitations of this type of materials in view of their application as electron sources.
For this purpose, we present an overview on the electronic properties of \ce{CsK2Sb}, including the analysis of the fundamental gap and of the character of valence and conduction bands.
We study the optical response in terms of the real and imaginary part of the dielectric function, discussing the role of the electron-hole interaction in the absorption and analyzing the character of the excitations.
Thanks to the capability of our approach to access core excitations on the same footing as the optical ones, we additionally investigate the x-ray absorption spectrum of \ce{CsK2Sb} from the Cs $L_3$-edge. 
This analysis offers additional information about the electronic structure of the system.
We finally discuss the results of this \textit{ab initio} many-body study in the context of the photocathode technology for particle accelerators with the goal of establishing a robust connection with the methods, the concepts, and the quantities typically obtained and analyzed in theoretical condensed-matter physics. 

\begin{figure}
\includegraphics[width=.5\textwidth]{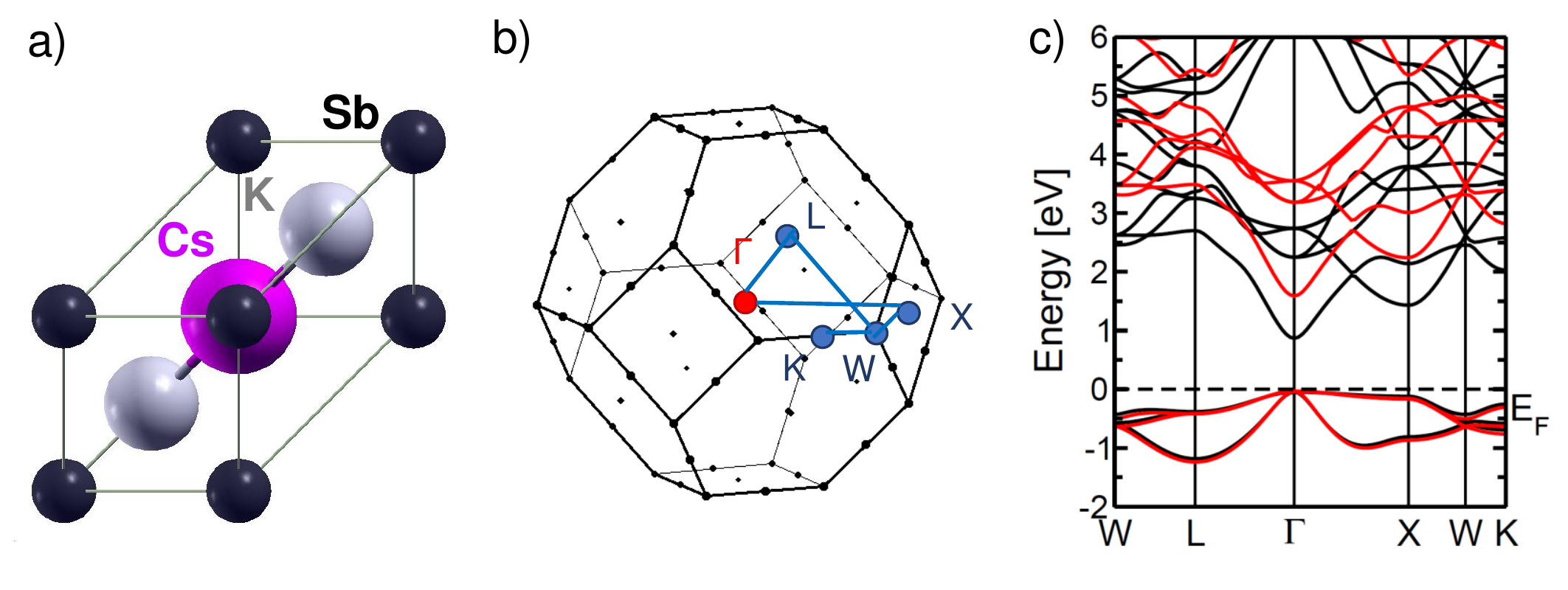}
\caption{(a) Ball-and-stick model of the unit cell of \ce{CsK2Sb} with Cs atom in magenta, K atoms in grey, and Sb atoms in black. (b) Brillouin zone of \ce{CsK2Sb} with high-symmetry points and the path connecting them highlighted. (c) Band structure of \ce{CsK2Sb} computed from DFT (black lines) and within the single-shot $G_0W_0$ approximation (red lines). The Fermi energy ($E_F$) is set to zero at valence-band top.}
\label{fig:structure}
\end{figure} 
%

\section{Cesium potassium antimonide: A promising semiconducting material for bright electron sources}
\label{section:system}

Photocathode materials commonly used for accelerator applications include metals and semiconductors, working within the ultra-violet (UV) and visible range of emitted light. 
Due to their high QE in the visible region, multi-alkali antimonides, such as \ce{CsK2Sb} (crystal structure shown in Fig.~\ref{fig:structure}a), have been used since the 1960s as photon detectors, \textit{e.g.}, for Cherenkov light in cosmic rays as well as for particle detection~\cite{somm56iretns}.
The development of this class of photocathode materials was additionally driven by their robustness, low dark current, and short response time~\cite{musu+18nimpra}.
Cs-K-Sb photocathodes are particularly attractive for high current and low emittance photoinjection. 
Their small band gap ($E_{gap}$) and low electron affinity (EA), both of the order of 1 eV~\cite{nath-mee67ije,ghos-varm78jap}, allow for photoemission close to threshold in the near-infrared (IR) or visible range, whilst at the same time sustaining a high QE.
The sum of the band gap and of the EA corresponds to the \textit{minimum energy for photoemission}, $E_{min}=E_{gap} + EA$, which in turn enters the expression of the intrinsic emittance ($\varepsilon_{ps}$) from the photocathode as follows:
\begin{equation}
\varepsilon_{ps}=\sigma_{x} \sqrt{\frac{h\nu - E_{min}}{3mc^{2}}}.
\label{eq:intrinsic_emittance}
\end{equation}
In Eq.~\eqref{eq:intrinsic_emittance}, $\sigma_{x}$ is the root-mean-square beam width and $h\nu$ is the energy of the incident photon. 
In the denominator we find universal constants such as the electron mass and the speed of light.
The intrinsic emittance from the emission process usually limits the brightness of the electron source.
The dependence of the $\varepsilon_{ps}$ on the square root of the minimum energy for photoemission suggests that by minimizing this quantity also the emittance decreases accordingly. 
It was indeed demonstrated experimentally that Cs-K-Sb photocathodes exhibit such characteristics~\cite{vecc+11apl}.
We note, however, that the relation given in Eq.~\eqref{eq:intrinsic_emittance} has been initially derived for metals and is based on a simplified model for the emission from discrete energy levels~\cite{berg-spic64pr,dowe-schm09prab}. 
While the qualitative picture suggested by this formula can be extended also to semiconductors, a quantitative description of the intrinsic emittance (and analogously of the QE) requires specific knowledge of the microscopic features of the material, which differ substantially between metals and semiconductors.
This is the type of information that the quantum-mechanical description of DFT and MBPT can provide, as demonstrated in the next sections.

\section{Theoretical Background and Computational Details}
\label{section:theory}

Density-functional theory lays its foundation onto the Hohenberg-Kohn theorems~\cite{hohe-kohn64pr} and is implemented according to the Kohn-Sham (KS) scheme~\cite{kohn-sham65pr}, which consists of mapping the many-electron problem into a set of independent-particle equations for the electronic wave-functions of each electron $i$ in the system:
\begin{equation}
\left[-\dfrac{\nabla^2}{2} + v_s(\mathbf{r}) \right] \varphi_i(\mathbf{r}) = \epsilon^{KS}_i \varphi_i(\mathbf{r}),
\label{eq:KS}
\end{equation}
where $\epsilon^{KS}_i$ is the KS energy per particle.
Eq.~\eqref{eq:KS} is expressed in atomic units, which are adopted from now on.
On the left-hand side of Eq.~\eqref{eq:KS}, in addition to the kinetic-energy operator, we find the effective potential per particle $v_s(\mathbf{r})$, which consists of the sum of three terms: $v_s(\mathbf{r}) = v_{ext}(\mathbf{r}) + v_H(\mathbf{r}) + v_{xc}(\mathbf{r})$. 
The external potential $v_{ext}$ includes the interaction between the negatively-charged electrons and the positively-charged nuclei.
The Hartree potential $v_H(\mathbf{r})$ accounts for the (mean-field) Coulomb between the electrons and $v_{xc}(\mathbf{r})$ is the exchange-correlation (xc) potential.
Since the exact form of $v_{xc}(\mathbf{r})$ is unknown, this term in Eq.~\eqref{eq:KS} has to be approximated. 
A reasonable approximation consists in treating exchange and correlation effects as in the homogeneous electron gas.
This is given by the local-density approximation (LDA).
In this work, we adopt the generalized gradient approximation (GGA) as implemented in the Perdew-Burke-Ernzerhof parameterization~\cite{pbe} and rely on MBPT for the description of excited-state properties, as detailed below.
A relevant aspect in the solution of the KS equations~\eqref{eq:KS} is the choice of the basis set.
Here, we work in the framework of the linearized augmented planewave plus local-orbital (LAPW+lo) method, which allows for an explicit treatment of core electrons in addition to valence and conduction ones.
An extensive description of this method is provided in Ref.~\cite{gula+14jpcm}.

Electronic properties and excitations are computed from MBPT with the DFT ground state as starting point.
The $GW$ approximation~\cite{hedi65pr} is adopted in the single-shot perturbative approach $G_0W_0$ on top of the solutions of the KS equations~\cite{hybe-loui85prl}. 
In this way one can compute the electronic self-energy $\Sigma$ that yields the quasi-particle (QP) energies of each electronic band $\epsilon_{i\mathbf{k}}$:
\begin{equation}
\epsilon_{i\mathbf{k}}^{QP} = \epsilon_{i\mathbf{k}}^{KS} + Z_{i\mathbf{k}} \left[ \Re \Sigma_{i\mathbf{k}}(\epsilon_{i\mathbf{k}}^{KS}) - V^{xc}_{i\mathbf{k}} \right],
\label{eq:QP}
\end{equation}
where $Z_{i\mathbf{k}}$ is the renormalization factor accounting for the energy-dependence of the self-energy.
For the derivation of Eq.~\eqref{eq:QP} and additional details we refer for review to Refs.~\cite{arya-gurn98rpp,onid+02rmp}.

Optical and core excitations are computed from the solution of the BSE~\cite{salp-beth51pr}, which is the equation of motion for the two-particle electron-hole Green's function~\cite{stri88rnc}.
The matrix form of the corresponding Schr\"odinger equation in the Tamm-Dancoff approximation reads:
\begin{equation}
\sum_{o'u'\mathbf{k'}} \hat{H}^{BSE}_{ou\mathbf{k},o'u'\mathbf{k'}} A^{\lambda}_{o'u'\mathbf{k'}} = E^{\lambda} A^{\lambda}_{ou\mathbf{k}} ,
\label{eq:BSE}
\end{equation}
where $o$ and $u$ label initial occupied and final unoccupied states, respectively.
In optical excitations initial states correspond to valence bands while upon x-ray absorption core electrons are excited to the conduction bands.
In the latter case, the \textbf{k}-dependence of the initial states drops (see for details Ref.~\cite{vorw+17prb}), although in Eqs.~\eqref{eq:Hdir} and~\eqref{eq:Hx} below it is retained for sake of generality.
In spin-unpolarized systems, the effective two-particle Hamiltonian $\hat{H}^{BSE}$ is expressed by the sum of three terms:
\begin{equation}
\hat{H}^{BSE} = \hat{H}^{diag} + \hat{H}^{dir} + 2 \hat{H}^x.
\label{eq:H_BSE}
\end{equation}
The \textit{diagonal} term $\hat{H}^{diag}$ accounts for single quasi-particle transitions, neglecting any Coulomb interaction between electron and hole, which is instead included in the two remaining terms.
The attractive interaction between the positively-charged hole and the negatively-charge electron is given by $\hat{H}^{dir}$:
\begin{equation}
\hat{H}^{dir} \! \! = \! - \! \! \int \! \! \! d^3\mathbf{r} \! \!  \int \! \! d^3\mathbf{r}' \phi_{o\mathbf{k}} (\mathbf{r}) \phi^*_{u\mathbf{k}} (\mathbf{r}') W(\mathbf{r},\mathbf{r}') \phi^*_{o'\mathbf{k}'} (\mathbf{r}) \phi_{u'\mathbf{k}'} (\mathbf{r}').
\label{eq:Hdir}
\end{equation}
This integral contains the screened Coulomb interaction $W = \varepsilon^{-1} v$, where $\varepsilon$ is the dielectric tensor of the system here computed within the random-phase approximation (RPA).
The third term in Eq.~\eqref{eq:H_BSE}, $\hat{H}^x$, represents the exchange interaction between the electron and the hole:
\begin{equation}
\hat{H}^x = \int d^3\mathbf{r} \int d^3\mathbf{r}' \phi_{o\mathbf{k}} (\mathbf{r}) \phi^*_{u\mathbf{k}} (\mathbf{r}) \bar{v}(\mathbf{r},\mathbf{r}') \phi^*_{o'\mathbf{k}'} (\mathbf{r}') \phi_{u'\mathbf{k}'} (\mathbf{r}'),
\label{eq:Hx}
\end{equation}
where $\bar{v}$ is the short-range part of the bare Coulomb potential accounting for local-field effects (LFE).
These effects are particularly relevant in case of inhomogeneous materials, where the impinging electromagnetic field can trigger charge fluctuations on the interatomic scale and thus generate internal induced microscopic fields~\cite{wise63pr,aspn82ajp,onid+02rmp}.
In the cubic \ce{CsK2Sb} crystal studied here, LFE are not expected to play a prominent role.

The BSE eigenvalues $E^{\lambda}$ (Eq. \ref{eq:H_BSE}) yield \textit{singlet} excitation energies.
The exciton binding energy ($E_b$) of bound electron-hole pairs is often defined as the difference between excitation energies and the minimum direct band gap: $E_b = E^{\lambda} - E_{gap}$.
This definition, however, does not apply to bound excitons above the absorption onset which can appear in complex materials, interfaces, and nanostructures (see, \textit{e.g.}, Ref.~\cite{aggo+17jpcl}).
For this reason, we adopt an alternative and more general definition of $E_b$ as the difference between the excitation energy computed by solving the BSE and obtained within the independent QP approximation (IQPA), \textit{i.e.}, retaining only $\hat{H}^{diag}$ in the Hamiltonian in Eq.~\eqref{eq:H_BSE}.
For bright excitons below the absorption onset the two definitions coincide.

The eigenvectors $A^{\lambda}$ appear in the expression of the transition coefficients between each pair of occupied and unoccupied states:
\begin{equation}
\mathbf{t}^{\lambda}= \sum_{ou\mathbf{k}} A^{\lambda}_{ou\mathbf{k}} \frac{\langle o\mathbf{k}|\widehat{\mathbf{p}}|u\mathbf{k}\rangle}{\varepsilon^{QP}_{u\mathbf{k}} - \varepsilon^{QP}_{o\mathbf{k}}} .
\label{eq:t}
\end{equation}
In the numerator of Eq.~\eqref{eq:t} we find the momentum matrix elements (equivalent to the transition-dipole moments) between initial and final states.
In the denominator the difference between QP energies of each pair of initial and final states appear.
For core excitations $\varepsilon^{QP}_{o\mathbf{k}} = \varepsilon^{KS}_o$. 
In this case, the \textbf{k}-dependence vanishes and the QP correction is replaced by a scissors operator, due to the limitation of the $G_0W_0$ approximation to quantitatively reproduce core-level energies~\cite{vorw+17prb}.
The transition coefficients from Eq.~\eqref{eq:t} enter the expression of the imaginary part of the macroscopic dielectric function representing the optical absorption of the material:
\begin{equation}
\Im\epsilon_M = \dfrac{8\pi^2}{\Omega} \sum_{\lambda} |\mathbf{t}^{\lambda}|^2 \delta(\omega - E^{\lambda}) .
\label{eq:ImeM}
\end{equation}
In Eq.~\eqref{eq:ImeM}, $\Omega$ is the unit cell volume and $\omega$ the energy of the impinging photon.
The real part of $\epsilon_M$ is related to the imaginary part by the Kramers-Kronig relations.
The loss function is given by $\mathcal{L} = -\Im\epsilon_M^{-1}$.

The BSE eigenvectors $A^{\lambda}$ of Eq.~\eqref{eq:BSE} contain information about the character and composition of the electron-hole pairs.
They appear in the expression of the two-particle excitonic wave-function
\begin{equation}
\Psi^{\lambda}(\mathbf{r}_{e},\mathbf{r}_{h}) = \sum_{ou \mathbf{k}} A^{\lambda}_{ou\mathbf{k}}\phi_{u\mathbf{k}}(\mathbf{r}_{e})\phi^{*}_{o\mathbf{k}}(\mathbf{r}_{h}),
\label{eq:psi}
\end{equation}
which is a six-dimensional quantity depending explicitly on the electron and the hole coordinates. 
In reciprocal space, we define the \textit{weights} of the hole and the electron as:
\begin{equation}
w^{\lambda}_{o\mathbf{k}} = \sum_u |A^{\lambda}_{ou\mathbf{k}}|^2
\label{eq:wh}
\end{equation}
and 
\begin{equation}
w^{\lambda}_{u\mathbf{k}} = \sum_o |A^{\lambda}_{ou\mathbf{k}}|^2,
\label{eq:we}
\end{equation}
respectively.
Summations in Eqs.~\eqref{eq:wh} and~\eqref{eq:we} run over the range of occupied and unoccupied states included in the solution of Eq.~\eqref{eq:BSE}.
Details about the implementation of MBPT within the LAPW+lo formalism can be found in Refs.~\cite{pusc-ambr02prb,Sagmeister2009,nabo+16prb,vorw+17prb}.

All calculations presented in this work are performed with the \texttt{exciting} code~\cite{gula+14jpcm}.
The muffin-tin (MT) radii of all the atomic species involved (Cs, K, and Sb) are set to 1.65 bohr and a plane-wave basis-set cutoff $R_{\textrm{MT}}G_{\textrm{max}} = 8.0$ is employed.
For ground-state calculations the Brillouin zone (BZ) is sampled using a homogeneous cubic \textbf{k}-mesh with 8 points in each direction, corresponding to overall 29 points considering the symmetry operations.
To calculate the quasi-particle band-structure within the $G_0W_0$ approximation, a 4$\times$4$\times$4 \textbf{k}-mesh is adopted without exploiting symmetries, for a total of 64 \textbf{k}-points.
In the BSE calculations for optical and x-ray absorption spectra, a $\Gamma$-shifted \textbf{k}-mesh with 8 and 6 points in each direction is adopted, respectively. 
To compute the screened Coulomb interaction within the RPA, 200 empty states are included.
LFE are accounted for by employing about 59 $|\mathbf{G}+\mathbf{q}|$ vectors in the BSE calculations for optical spectra and about 300 in those for x-ray spectroscopy.
Unit cells and BZs are visualized with the XCrysDen software~\cite{xcrysden}.

\section{Results}
\label{section:results}
\subsection{Electronic properties}
\begin{figure}
\center
\includegraphics[width=.5\textwidth]{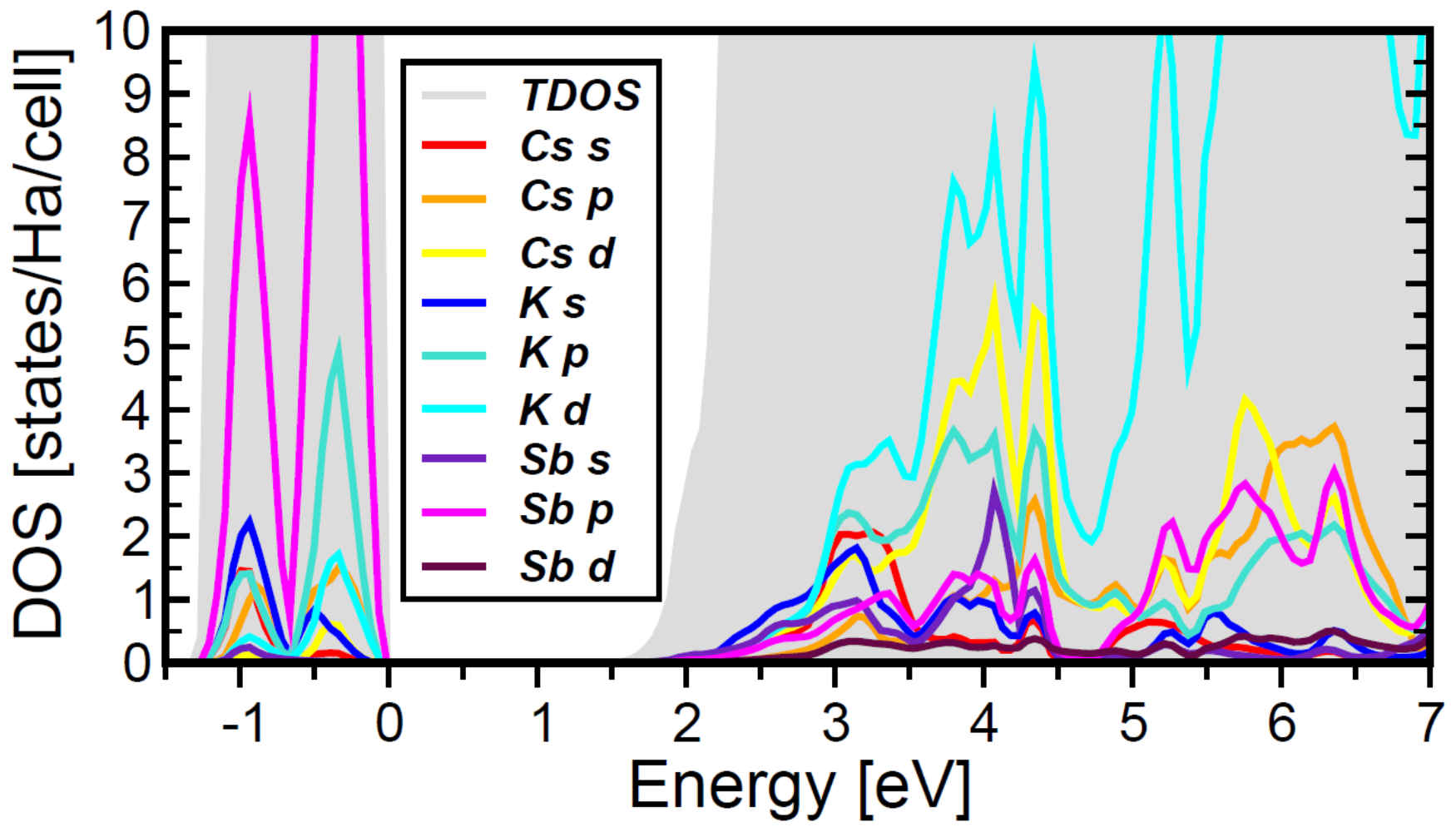}
\caption{Total density of states (TDOS, grey area) of \ce{CsK2Sb}, including the QP correction as a rigid shift. Atom-projected electronic contributions depicted in color. The Fermi energy is set to zero at valence-band top.}
\label{fig:pdos}
\end{figure} 
We start our analysis by inspecting the electronic structure of \ce{CsK2Sb}, which is a face-centered cubic crystal including four atoms in its unit cell.
The lattice parameter $a$=8.76 \AA{} has been obtained by performing volume optimization and fitting the results with the Birch-Murnaghan equation of state~\cite{birch,murnaghan}.
This result is in agreement with the experimental value reported in Ref.~\cite{somm63apl} and with previous DFT predictions~\cite{kala+10jpcs,murt+16bms}, obtained, however, utilizing a different parameterization of the GGA xc functional.
As shown in Fig.~\ref{fig:structure}a, the Sb atom is located at the origin (0.0; 0.0; 0.0), while the Cs is in the center at (0.5; 0.5; 0.5) and the two K are at ($\pm$0.25; $\pm$0.25; $\pm$0.25)~\cite{ette-degr02prb}.
The BZ is depicted in Fig.~\ref{fig:structure}b with the $\Gamma$-point highlighted in red and the other high-symmetry points used in the band-structure plot in Fig.~\ref{fig:structure}c marked in blue.
There, the result obtained from $G_0W_0$ (red lines) is compared to the one from DFT (black lines).
The two approaches yield very similar band structures: The valence bands coincide and the direct band gap is at $\Gamma$.
Dispersion and effective masses in the conduction region are almost identical.
The most striking difference, which is not surprising at all, is the size of the band gap.
DFT gives a value of 0.92 eV which is drastic underestimated compared to the $G_0W_0$ one of 1.62 eV, including the QP correction.
Previously computed values of the band gap of \ce{CsK2Sb} exhibit a large variability, depending on the adopted approximations for the xc functional and the basis set.
Results from the literature range from 0.58 eV~\cite{ette-degr02prb}, to 0.80 eV~\cite{murt+16bms} up to 1.13 eV~\cite{kala+10jpcs}.
In Ref.~\cite{murt+16bms} additional estimated of 1.58 eV and 1.99 eV are obtained with the hybrid xc functionals proposed by Engel and Vosko~\cite{enge-vosk93prb} and Tran and Blaha~\cite{tran-blah09prl}, respectively.
Both values are in line or slightly exceed the $G_0W_0$ result presented here.
The few experimental references available for \ce{CsK2Sb} date back several decades ago. 
The early study by Nathan and Mee suggests for this material a band gap of 1.0 eV and an electron affinity 1.1 eV~\cite{nath-mee67ije}.
Photoconductivity measurements by Ghosh and Varma, yielded a similar gap of 1.2 eV~\cite{ghos-varm78jap}.
Our $G_0W_0$ result of 1.62 eV, obtained for the ideal single crystal, infinitely periodic, and free from any defect or impurity at zero temperature, overestimates these values.
However, it is hard to perform a comparison between theory and experiment in the absence of more recent and robust references. 
For this purpose, new measurements on high-quality samples grown under controlled conditions are required. 

\begin{figure*}
\center
\includegraphics[width=.9\textwidth]{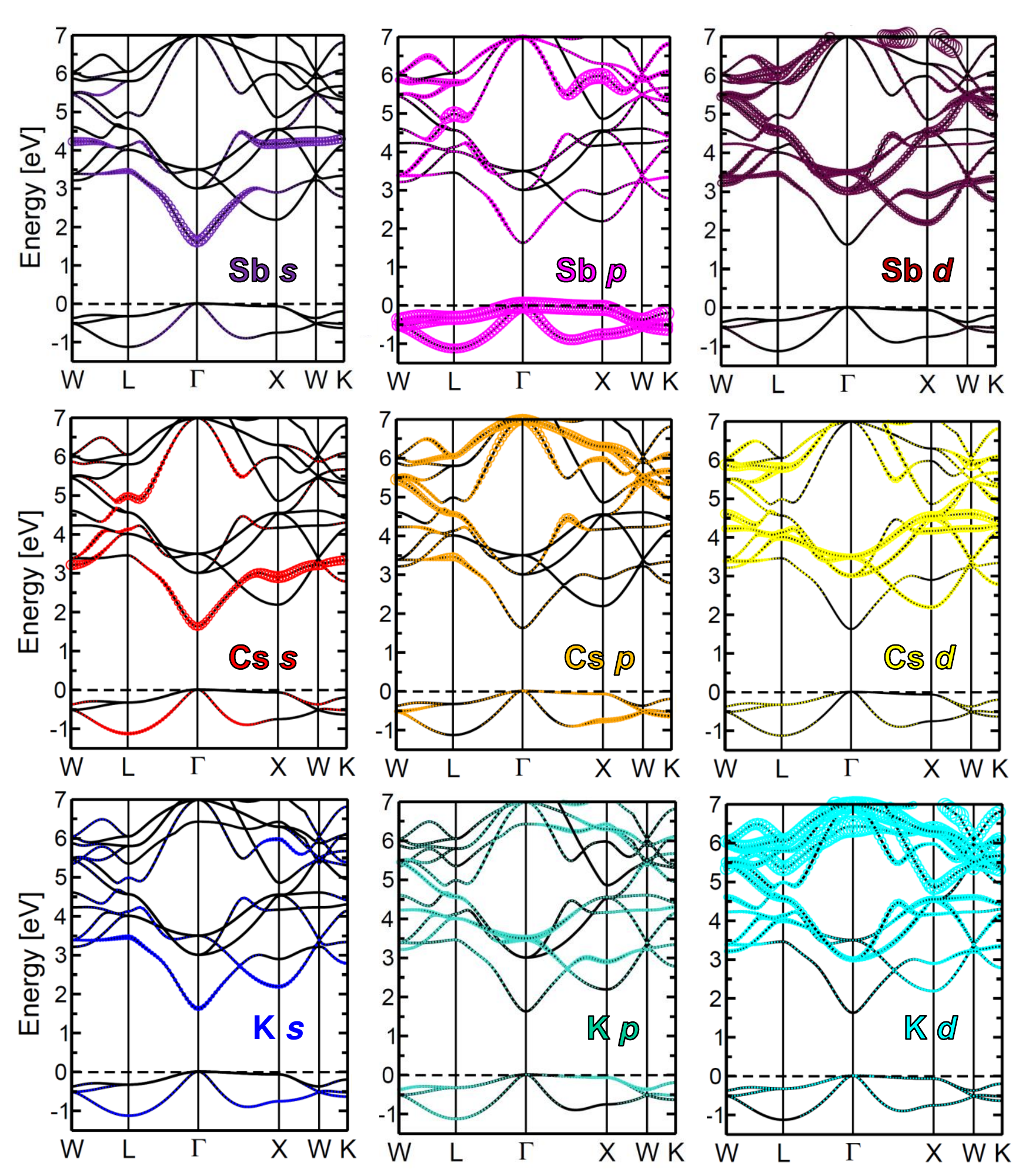}
\caption{Band structure plots of \ce{CsK2Sb}, including the QP correction as a rigid shift, showing the atom-projected character of the valence and conduction bands, quantified by the size of the colored circles. The contributions of Sb $d$-states are magnified by a factor 10 compared to all the others.}
\label{fig:bands}
\end{figure*} 
The character of the electronic bands is a crucial piece of information to identify the selection rules for allowed optical transitions, which are the first step of the photoemission process. 
For this purpose, we consider both the atom-projected density of states (DOS), shown in Fig.~\ref{fig:pdos}, and the QP band structures of \ce{CsK2Sb} with the atom-projected contributions (Fig.~\ref{fig:bands}).
One of the peculiarity of \ce{CsK2Sb} and, in general, of all alkali antimonides, is the limited number of valence states in the vicinity of the band gap (see Figs.~\ref{fig:structure}c and~\ref{fig:bands}).
Lower-energy bands appear with almost flat dispersion at about 7.5 eV below the Fermi energy (not shown), as discussed in previous works~\cite{ette-degr02prb,kala+10jpcs,murt+16bms}. 
In line with the results of Refs.~\cite{ette-degr02prb,kala+10jpcs,murt+16bms}, obtained only at the DFT level, the three highest-occupied bands exhibit a clear $p$-character with contributions from the Sb atoms and, to less extent, from K and Cs $p$-states (Figs.~\ref{fig:pdos} and~\ref{fig:bands}). 
Potassium $s$-electrons also contribute, as shown in the DOS in Fig.~\ref{fig:pdos}.
The conduction region, although strongly hybridized, is dominated by $d$-electrons, with the exception of the lowest-unoccupied band which has predominant $s$-character, as also suggested by its parabolic shape.
The conduction-band minimum at $\Gamma$ has hybridized Cs-Sb character, while along the X-W-K path most relevant contributions are given by Cs $s$ electrons.
Above 3.5 eV in the energy scale of Fig.~\ref{fig:bands}, Cs and K $d$-electrons dominate.
Sb $d$-states hybridized with those bands also appear but their contribution is one order of magnitude smaller (the size of the circles in the corresponding plot in Fig.~\ref{fig:bands} is magnified by a factor 10).

\subsection{Optical properties}
\begin{figure*}
\includegraphics[width=\textwidth]{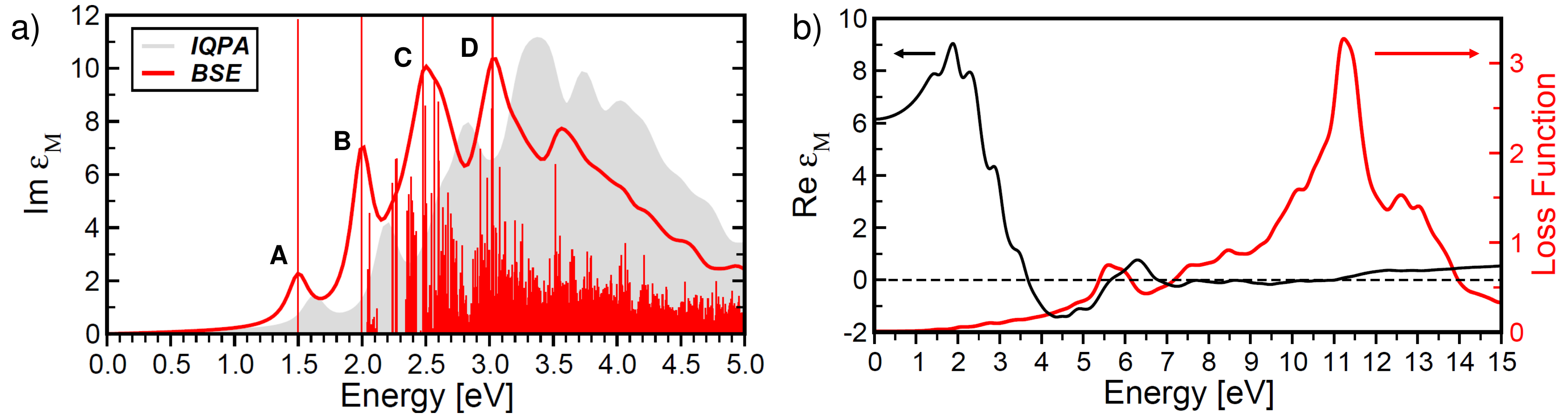}
\caption{a) Imaginary part of the macroscopic dielectric function of \ce{CsK2Sb} including excitonic effects (BSE, red line). Vertical bars indicate position and relative intensity of the solutions of the BSE. The spectra include a Lorentzian broadening of 100 meV mimicking the excitation lifetime. The first four peaks are labeled as A, B, C, and D respectively. For comparison, the spectrum without excitons, computed in the so-called independent quasi-particle approximation (IQPA, grey area), is also shown. b) Real part of the macroscopic dielectric function (black) and loss function at $\mathbf{q} \rightarrow 0$ (red) computed from the BSE. A dashed line marks the zero in the scale of Re $\varepsilon_M$ as a guide for the eyes. A Lorentzian broadening of 200 meV is applied.}
\label{fig:bse}
\end{figure*} 
We now turn to the optical properties of \ce{CsK2Sb}.
The dielectric function of this material was previously reported in Refs.~\cite{kala+10jpcs,murt+16bms} for the ideal crystal and in Ref.~\cite{kala+10jpcs1} for the crystal under pressure.
In these works, however, electron-hole correlation effects were not included.
In Fig.~\ref{fig:bse}a we show the absorption spectrum of \ce{CsK2Sb} computed from the solution of the BSE.
Since Eq.~\eqref{eq:BSE} is solved by full diagonalization, we have access to the eigenvalues $E^{\lambda}$, representing the excitation energies marked by vertical lines in Fig.~\ref{fig:bse}a.
In the same graph, we also show the IQPA spectrum which is analogous to the ones reported in Refs.~\cite{kala+10jpcs,murt+16bms}.
The only difference is that here the initial and final states of the optical transitions include the QP correction computed from $G_0W_0$.

\begin{figure*}
\center
\includegraphics[width=.95\textwidth]{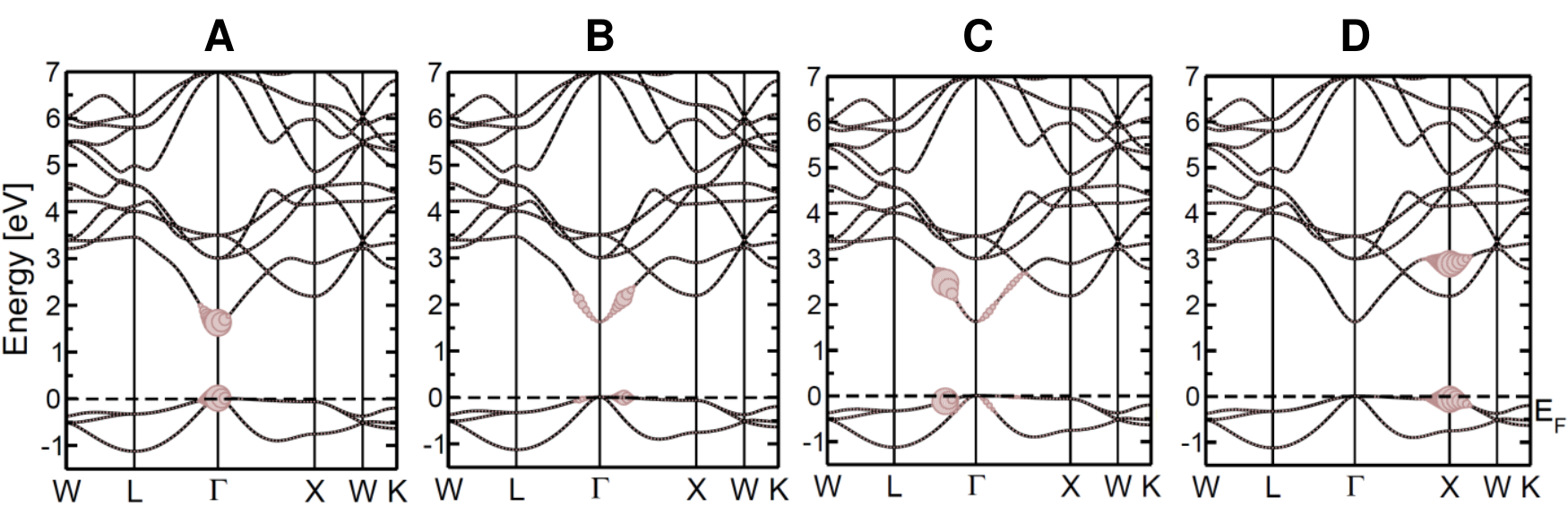}
\caption{Band-structure contribution (circles, see Eq.~\ref{eq:wh} and~\ref{eq:we}), of the excitons labeled in the optical absorption spectrum of Fig.~\ref{fig:bse}a). The QP correction is included as a rigid shift and the Fermi energy ($E_F$) is set to zero at valence-band top.}
\label{fig:excitons}
\end{figure*} 

The absorption onset of \ce{CsK2Sb} lies in the visible region at 1.5 eV and is given by a relatively weak peak, labeled A, formed by an exciton which is three-fold degenerate due to the cubic symmetry of the crystal. 
It stems from transitions between the valence-band maximum (VBM) and the conduction-band minimum (CBm) around the $\Gamma$-point (see Fig.~\ref{fig:excitons}) and has a binding energy of 120 meV.
The second peak, labeled B, also lies in the visible region at approximately 2 eV.
It is more intense than A and is formed by a manifold of excitations, which are again three-fold degenerate by crystal symmetry. 
The most intense excitations giving rise to B emerge from interband transitions in the vicinity of the $\Gamma$-point, along the $\Gamma$-L and the $\Gamma$-X paths (Fig.~\ref{fig:excitons}).
The electron-hole interaction acts on these excitation by enhancing their oscillator strength compared to the IQPA spectrum.
Moreover, albeit peak B lies above the absorption onset, and, as such, the excitations forming it are conventionally not associated to bound excitons, their binding energy can be nonetheless estimated by comparing the position of this peak with its IQPA counterpart (see Refs.~\cite{aggo+17jpcl,aggo+18prb}).
The resulting binding energy is about 100 meV.
The bound nature of these excitations is further indicated by their separation in energy from the continuum of excited states starting with peak C at approximately 2.5 eV.
Also this peak is formed by a multitude of excitations.
In Fig.~\ref{fig:excitons} we report the band contributions to the most intense one, which corresponds again to a transition between the highest-occupied band and the lowest-unoccupied one but in this case along the $\Gamma$-L path.
Also for the excitations forming the peak C the binding energies can be estimated in comparison with the IQPA spectrum and are of the order of a few hundred meV. 
The oscillator strength of peak C is enhanced when the electron-hole interaction is taken into account.
A fourth peak, D, appears at the edge of the visible region at 3 eV.
It is also red-shifted compared to its IQPA counterpart by a few hundred meV and, while formed by hundreds of BSE solutions, it is dominated by a few intense excitations.
In Fig.~\ref{fig:excitons}, right panel, the composition of one of these excitations is shown: The dominant transitions are localized in the vicinity of the X point in the BZ.
In the conduction region, target states belong to the second-lowest unoccupied band.
By looking at Fig.~\ref{fig:bands}, the Sb $p$-$d$ character of these higher transitions is apparent.
Conversely, lower-energy peaks, formed by transitions in the vicinity of the $\Gamma$-point, have a prominent Sb $p$-$s$ character.
At higher energies, in the UV region, we find a continuum of excitations with almost homogeneously distributed spectral weight.
Overall, the whole BSE spectrum plotted in Fig.~\ref{fig:bse}a undergoes a red-shift compared to the IQPA. 
As pointed out above, almost all peaks undergo a redistribution of their intensity, with the ones in the visible band gaining weight at the expenses of those in the UV band.
These features indicate the relevant effects of the electron-hole interaction, which acts also beyond the absorption onset. 

Interaction with higher-energy photons gives rise to collective electronic oscillations, which are commonly identified as plasmons.
Plasmonic excitations appear in the loss function of \ce{CsK2Sb} at about 10 eV (Fig.~\ref{fig:bse}b).
Relevant features in this spectroscopic quantity appear well above the visible region, namely at $\hbar \omega >$ 5 eV.
While local maxima appear between 5 and 9 eV, the most prominent peak in the loss function is at 11 eV.
At this frequency, a pole in the real part of $\epsilon_M$ confirms the plasmonic nature of this excitation.
In the low-energy region, instead, local oscillations in the real part of $\epsilon_M$ correspond to the absorption peak positions in the imaginary part (see Fig.~\ref{fig:bse}a). 

\subsection{Core excitations from the Cs $L_3$ edge}
\begin{figure}
\center
\includegraphics[width=.5\textwidth]{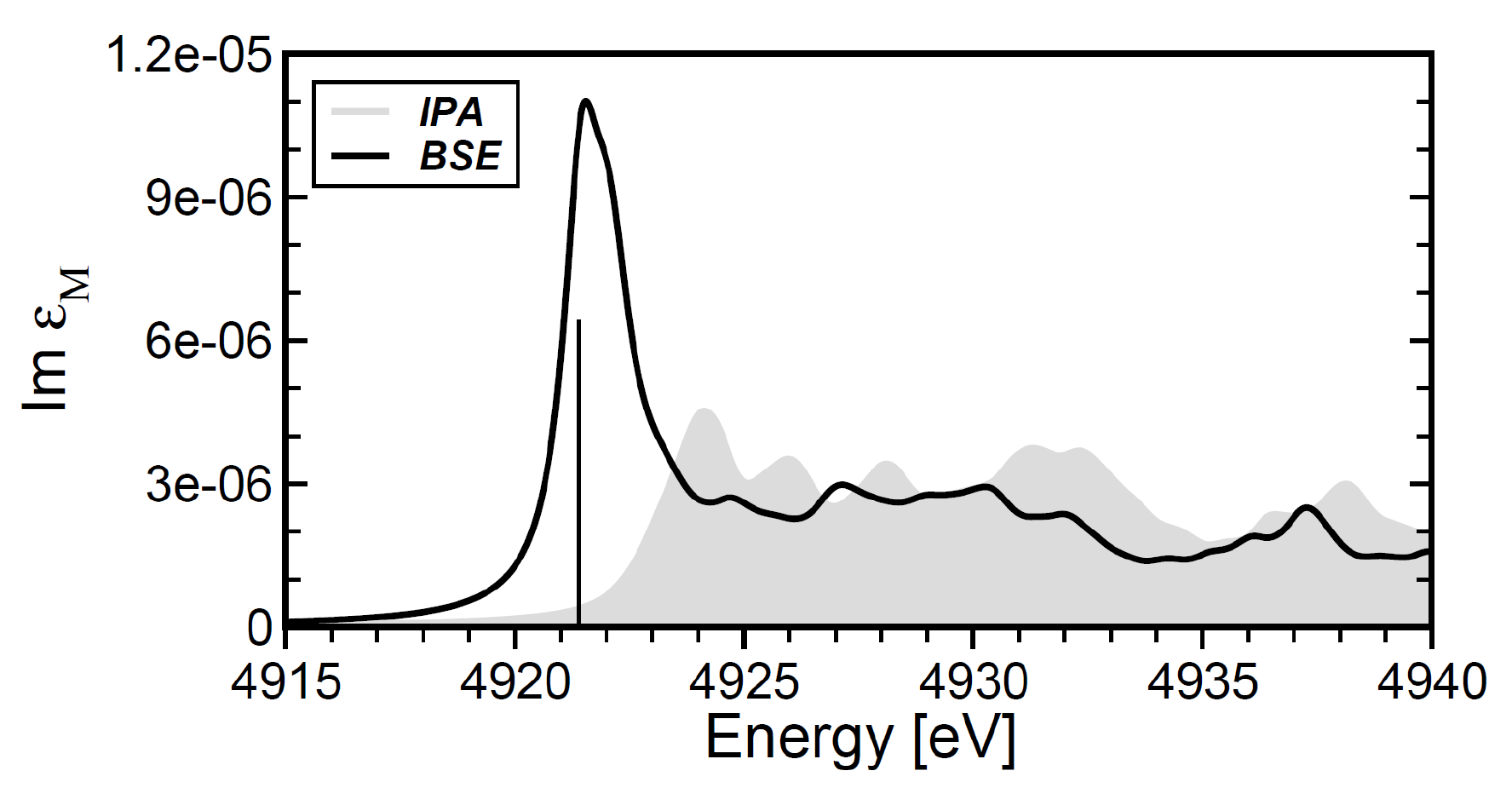}
\caption{X-ray absorption spectrum of \ce{CsK2Sb} from the Cs $L_3$-edge given by the imaginary part of the macroscopic dielectric function computed from the solution of the Bethe-Salpeter equation (BSE, black line). The vertical line marks the energy of the first excitons analyzed in Fig.~\ref{fig:excitons-xas}. The result obtained in the independent-particle approximation (IPA) neglecting excitonic effects is also shown for comparison (grey area). A Lorentzian broadening of 500 meV is applied to both spectra to mimic the excitation lifetime.}
\label{fig:xas}
\end{figure} 

To complete the spectroscopic characterization of \ce{CsK2Sb} we turn to core excitations, focusing on transitions from the Cs $L_3$ edge. 
According to the selection rules for the quantum number $l$, target levels are Cs $s$- and $d$-states, which indeed dominate the bottom of the conduction region (see Figs.~\ref{fig:pdos} and ~\ref{fig:bands}).
The absorption spectrum is computed from the solution of the BSE  wit the set of initial states restricted to Cs $2p_{3/2}$ electrons.
It is possible to decouple these transitions from those from the Cs $2p_{1/2}$ states due to their large spin-orbit splitting of a few hundreds eV.
For comparison, we plot in Fig.~\ref{fig:xas} also the spectrum computed without the electron-hole interaction, in the independent-particle approximation (IPA).
In this case, transitions are calculated directly from the KS states, not including the QP correction.
The $GW$ approximation is in fact unable to account for the QP correction of the extremely localized core states, which are typically underestimated by 10$\%$ of their energy in DFT.
Adding the QP correction from $G_0W_0$ to the conduction bands, which is of the order of a few hundred meV, would thus not yield the correct absorption onset, which is indeed underestimated by roughly the same amount~\cite{vorw+17prb}. 
Also the spectrum shown in Fig.~\ref{fig:xas} suffers from this drawback. 
However, in the absence of experimental references, which typically guide the choice of an appropriate scissors operator (see, \textit{e.g.}, Refs.~\cite{cocc-drax15prb,cocc+16prb,vorw+18jpcl}), we decide to show the result of our calculation without the includion of any rigid shift.
From the difference between the KS core energy of Cs $2p_{1/2}$ electrons and reference values~\cite{bear-burr67rmp} and based on our experience~\cite{cocc-drax15prb,cocc+16prb,vorw+18jpcl}, we expect the absorption onset in Fig.~\ref{fig:xas} to be underestimated by a few tens of eV.

\begin{figure}
\center
\includegraphics[width=.5\textwidth]{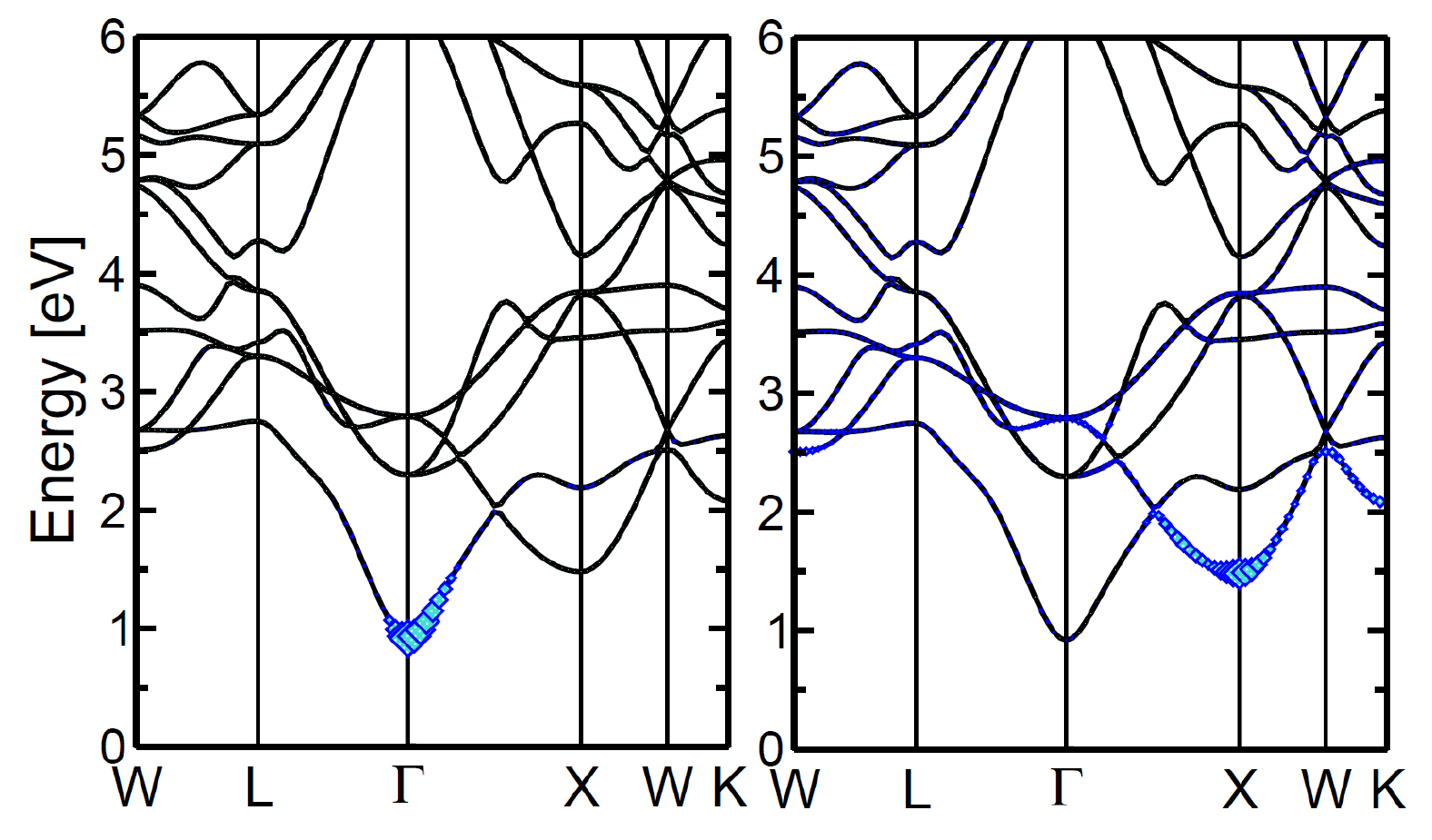}
\caption{Conduction-band contribution (blue diamonds) of two (almost degenerate) excitons forming the sharp excitonic peak in the x-ray absorption spectrum of \ce{CsK2Sb} from the Cs $L_3$-edge, marked by the vertical bar in Fig.~\ref{fig:xas}. The origin corresponds to the valence-band top.}
\label{fig:excitons-xas}
\end{figure} 

The absorption onset of the BSE spectrum in Fig.~\ref{fig:xas} is characterized by a sharp maximum.
In absolute units, its intensity is several orders of magnitude smaller than the maxima in the optical absorption spectrum (Fig.~\ref{fig:bse}).
This difference is determined by the significantly smaller wave-function overlap between conduction bands and core states compared to the valence ones.
Concomitantly, the energy difference between the involved electronic levels is orders of magnitude larger.
The numerator of Eq.~\ref{eq:t} is therefore reduced while the denominator is increased.
The comparison between the BSE spectrum and its IPA counterpart highlights the strong excitonic character of this peak.
Overall, it is formed by 16 excitations with binding energies ranging between 150 and 200 meV.
These values are about twice as large as those in the visible region, which is not surprising, given the more localized character of x-ray excitations compared to the optical ones.
Here, excitonic effects manifest themselves mainly through the sizable redistribution of the spectral weight towards to the first peak.
The IPA spectrum, obtained neglecting the electron-hole interaction, exhibits peaks of almost equal intensity, from the absorption edge up to 15 eV above it.

The band contributions to the first core bound excitons highlighted in the spectrum in Fig.~\ref{fig:xas} reveal that transitions to both $s$- and $d$-states are involved (see Fig.~\ref{fig:excitons-xas}).
The lowest-energy excitation corresponds to transitions to the CBm at $\Gamma$ (Fig.~\ref{fig:excitons-xas}a), which has a predominant $s$-character (see Fig.~\ref{fig:bands}).
Higher-energy excited states comprised in the first sharp peak are instead due to transitions to the band minimum at X, having a distinct $d$-character.
Non-negligible contributions also come from higher-energy bands (Fig.~\ref{fig:excitons-xas}b).
The coexistence of these excitations with different character within the first absorption peak makes the analysis of x-ray spectra from the Cs $L_3$ edge particularly promising for characterizing different phases and chemical compositions of cesium antimonides. 

\section{Discussion}
\label{section:discussion}
The results presented above show the wealth of insight that first-principles many-body calculations can provide on the electronic structure of a material and on its excitations.
This information can be fruitfully exploited to model the characteristics of semiconducting photocathodes for particle accelerators. 
Band gap and electron affinity, the key parameters to estimate the intrinsic emittance of the cathode, can be reliably computed only beyond the mean-field approach provided by semi-local DFT, which is unable to access excited-state properties.
Adopting MBPT approaches also allows to overcome the large variability of the DFT results related to different basis sets and xc functionals.
Additional information about the electronic structure is given by the atomic character of the bands and their hybridization.
These properties are crucial to understand the selection rules of optical and core-level spectroscopy.
The optical absorption of \ce{CsK2Sb}, which spans the entire visible band, is characterized by excitonic effects also above the onset.
Their role is to red-shift the maxima by a few hundred meV and to redistribute the oscillator strength towards the lower-energy peaks. 
The electronic transitions involved are mainly between the uppermost valence bands and the lowest conduction bands, with different contributions arising from various regions of the BZ, depending on the excitation energy. 
All these facets demonstrate that a simplistic description of the absorption process cannot capture the complex behavior semiconductors.
We have investigated x-ray absorption spectra of \ce{CsK2Sb} from the Cs $L_3$-edge, showing that also in this regime excitonic effects are dominant.
In general, the ability of our first-principles many-body approach to access core excitations from any atom and any edge is essential for achieving a comprehensive characterization of the material.
Core spectroscopy represents in fact a powerful tool to probe the electronic structure of materials with atomic resolution, in order to detect the presence of defects, impurities, dislocations, and more.

In the context of electron sources for particle accelerators, the main quantities that determine the performance of a photocathode are the intrinsic emittance (Eq.~\ref{eq:intrinsic_emittance}) and the QE, which estimates the amount of current that can be extracted from the material for given laser energy and power.
Parameters like effective electron masses, absorption coefficients, and band gaps, which are typically extracted from \textit{ab initio} calculations, can be substantially improved going beyond DFT and adopting the many-body framework presented here. 
The actual challenge, however, remains how to combine these quantities to realistically predict the performance of real cathodes remains a challenge.
Photoemission is typically represented in a three-step picture initially designed for metallic cathodes~\cite{berg-spic64pr}, where the description of the whole process can be further simplified into a one-step model~\cite{kark+17prb}.
The extension of the three-step model to semiconducting electron sources requires appropriate extensions that take into account the specific mechanisms occurring in these materials.
For instance, scattering events that are particularly relevant in photoemission, as (in a semi-classical picture) they randomize the momentum distribution of the excited electrons, are mainly dominated by electron-phonon interactions in semiconductors.
These processes cannot be modeled without having access to the microscopic properties of the material.
Recent attempts to describe the transport of excited electrons towards the surface have been successfully carried out with Monte-Carlo simulations for alkali antimonides~\cite{xie+16prab,gupt+17jap}.
Dedicated \textit{ab initio} studies focused on these types of interactions would further improve the current description and understanding of the photoemission process.

Extensive first-principles many-body studies of semiconducting materials for photocathodes like the one presented here are expected to bring significant benefits to the entire field of particle accelerator physics. 
The computational screening of a large number of compounds would further allow to identify systems with most promising characteristics in terms of QE performance.
This information, combined with the knowledge of deposition parameters and other electron beam specifications, could contribute to tailor the properties of the cathode.
For instance, one limiting factor in current photoinjector technology is the drive laser capability in the UV range. 
High-gain lasing is typically achieved in the IR region and then converted to UV by frequency up-conversion \cite{panu-piot17apl}.
However, this process is not efficient. 
First-principles investigations could point to the necessary steps to tailor the electronic structure in order to shift the sensitivity to longer wavelengths, in view of ultimately designing and growing photocathodes operating in the IR range.

Further degrees of freedom in modeling materials for electron sources are surface effects, which have been the subject of an extensive computational work on metals~\cite{cami+16cms}.
Also semiconductors like alkali antimonides are characterized by highly reactive surfaces that are prone to contamination outside of ultra high vacuum conditions.
Novel solutions are currently being explored to improve the lifetime of pure surface states.
In a recent computational study based on DFT it was proposed to adopt protective monolayer coatings deposited on \ce{Cs3Sb} photocathodes, which are able to improved their lifetime by preventing surface contamination by residual gas~\cite{wang+18npj2dma}.  

\section{Summary and Conclusions}
\label{sec:conclusions}
In summary, we have presented a comprehensive first-principles study of the electronic and optical excitations of \ce{CsK2Sb}, a semiconducting photocathode material for particle accelerators. 
We have characterized the band structure and the absorption spectrum in the framework of many-body perturbation theory, demonstrating the relevant role played by electron-electron and electron-hole correlation effects. 
With an estimated band-gap of about 1.6 eV \ce{CsK2Sb} is confirmed to be a promising electron source.
It is absorbs visible light over the entire visible range where it exhibits excitonic effects that mainly consist in a slight redistribution of the oscillator strength towards low-energy maxima.
We have analyzed the excitations in terms of band contributions, showing that transitions between the higher occupied and the lowest-unoccupied bands dominate the absorption at visible wavelengths.
This study in the optical region has been complemented by x-ray absorption spectroscopy computed from the Cs $L_3$-edge.
The corresponding absorption edge is dominated by an intense excitonic peak stemming from transitions to unoccupied bands with both Cs $s$- and $d$-character. 

The predictive power of the adopted methodology and the physical insight gained in the analysis of our results suggest a promising application of this approach to model photoemission from electron sources for particle accelerators. 
If the parameters obtained in this framework can be inserted directly in the existing models, the main challenge for future and more accurate studies consists in the extension of the model towards the inclusion of quantum-mechanical and many-body effects.
This perspective can set the basis for an improved modeling of semiconducting photocathodes for particle accelerators and, concomitantly, extend the range of action of the methods of computational condensed-matter physics to systematically screen and investigate new materials for electron sources.


\end{document}